# Cs+ sputtered clusters from multi-walled carbon nanotubes and graphite

Shoaib Ahmad[a,c*], Sumera Javeed[b], Sumaira Zeeshan[b], Athar Naeem[a], Shahzad Saadat[a], Muhammad Yousuf[a], Muhammad Khaleel[a], Ahsan Mushtaq[a], Muhammad Shahnawz[a]

[a]Government College University (GCU), CASP, Church Road, 54000 Lahore, Pakistan
[b]Pakistan Institute of Nuclear Science and Technology, P O Nilore, Islamabad, Pakistan
[c]National Centre for Physics, Quaid-i-Azam University Campus, Shahdara Valley, Islamabad, 44000, Pakistan
[*]Email: sahmad.ncp@gmail.com

**Abstract**

Experiments with multi-walled carbon nanotubes (MWCNTs) and graphite as targets in a source of negative ions with cesium sputtering (SNICS) have shown that MWCNTs with nm radii and μm length can be compared with μm-size graphite grains to understand the irradiation effects that include the formation, sputtering of carbon clusters and the resulting structural changes. The simultaneous adsorption of Cs° on the surface and bombardment by energetic $Cs^+$ is shown to play its role in the cluster formation and sputtering of carbon atoms and clusters ($C_x$; $x \geq 1$) and the cesium-substituted carbon clusters ($CsC_x$) as anions. Sputtered species' qualitative and quantitative outputs are related to their respective structures. Structural changes are shown to occur in MWCNTs and seen in SEM micrographs. The individual identity of the heavily bombarded MWCNTs may have given way to the merged structures while effects on the structure of heavily irradiated graphite grains size needs to be further investigated.

## 1. Introduction

Multi-walled carbon nanotubes (MWCNTs) resemble graphite in having the graphene sheets that are cylindrically bent as coaxial cylindrical tubes. Inter-tube spacing is also similar to the inter-sheet spacing in graphite (0.334 nm). The main difference appears in the dimensional characteristics of MWCNTs as compared to graphite. The grains of graphite have typical dimensions of few μm, implying volumes of ~ μm$^3$ that contain ~$10^{11}$ -$10^{13}$ C atoms. MWCNTs have one dimension in the nanometer range i.e. the diameters while their length is in μm. A 10 nm diameter x 1 μm long MWCNT contain ~$10^8$ C atoms. Polycrystalline graphite containing the individual μm sized grains is an electrical conductor. The individual MWCNTs retain their structural identity even in a sample of compressed MWCNTs and are poor conductors as a compressed sample. The similarities and differences in their respective structural properties make MWCNTs and graphite as distinct and at the same time, comparable carbon structures for the study of irradiation induced effects. In this study we focus on the structural changes introduced by energetic heavy ion irradiation in MWCNTs and graphite with the help of mass spectrometric identification of the sputtered species that are emitted. The SEM micrographs taken before and after the irradiations help in understanding the similarities and differences in the two sp$^2$-bonded irradiated-carbon allotropes' structures.

Extensive studies of various types of irradiation effects have been done on graphite for being the highest sublimation temperature material and due to its crucial role in nuclear reactors. Irradiation-induced physical property changes in graphite due to neutrons and high energy charged particle bombardments were studied [1]. Studies of the carbon cluster emissions under energetic particle bombardments identify and relate to the nature of the irradiation damage. The nature and type of the irradiation-induced damage in graphite is well documented [2-6]. The emergence of the new allotropic forms of carbon i.e. single and multi-shelled fullerenes, single and multi-walled nanotubes over the last two decades has renewed interest in carbon's sp$^2$-bonded structures. Due to the maneuverability of producing carbon nanotubes with desired sizes (diameter and length), chirality etc, these have been exhaustively studied. Irradiation studies extending over the last two decades, of the carbon nanotubes (single and multi-walled) have been reviewed recently with emphasis on various applications [7-9]. The above mentioned resemblance and the subtle differences between MWCNTs and small size grains of graphite provide us the indicators to understand and modify the radiation induced physical properties. Investigations of the sputtered carbon species $C_x; x \geq 1$ that contain mono-, di-, tri-atomic and higher clusters from irradiated graphite have been conducted [2-6]. We have extended these studies to monitoring the sputtered $C_x$ from MWCNTs with the view to understand the nanotube fragmentation pattern and to see the effects of the accumulation of the sputtered species in the intra-nanotube space. Do the emitted constituents, after leaving the parent nanotubes' outer, irradiated-surfaces escape and get deposited on and around other nanotubes? We may further ask, whether the increased concentrations of sputtered $C_x$ in the intra-MWCNT space lead to the emergence of new structures that have varying degrees of $sp^2$ and $sp^3$ bonding? We believe that in this study we have some evidence that the heavily irradiated MWCNTs fragment and certain structural transformations occur. SEM micrographs provide that evidence. The nature and the dynamics of these have been discussed mostly with reference to graphite [2,3]. The present study looks into the similarities and differences amongst the sputtered species from the two allotropes of carbon.

## 2. Experimental
### 2.1. Source of negative ions with cesium sputtering (SNICS)

The MWCNT samples were prepared in 8mm diameter x 20mm long, cylindrical Al bullets by compressing the MWCNTs in 6 mm diameter holes that are 2.5 mm deep and subsequently annealed for four hours at 500° C. The bullets filled with MWCNT and reactor grade graphite were used as targets for the negative ion source with cesium sputtering (SNICS) that is mounted on the 2 MV Pelletron at GCU, Lahore. SNICS with the analyzing magnet and Faraday Cage provide the experimental set up for removing the target material at a desired rate by sputtering and monitoring the intensities of the sputtered species as a function of their respective mass to charge ratios (m/z) in atomic mass units (amu). SNICS operates with a hollow ionizer that is situated between the target and the extractor. Neutral cesium Cs° vapour is drawn into the source from an external cesium reservoir that is heated to ~100-110°C. Cs° loses its charge on the ionizer surface and convert into a positively charged ion $Cs^+$. It is attracted towards the negatively charged target containing the to-be-sputtered species. The energy and intensity of the cesium ions $Cs^+$ delivered by SNICS remain steady over extended periods

~ hours. A stable heavy ion source is crucial for the study of ions like $Cs^+$ for inducing fragmentations leading to the sputtering of clusters. A detailed profile of the sputtering by $Cs^+$ in cesium negative ion sources was highlighted by Middleton [10]. His study included the sputtering of molecular solids along with the metallic ones. From graphite large number of carbon monatomic, diatomic and higher clusters $C_x$ (x=1-10) and the cesium-substituted clusters $CsC_x$ (x≥1) were also shown to be among the sputtered species. The importance of the role of surface coverage by neutral $Cs°$ had been pointed out much earlier [11]. It is due to the high sputtering yields for most target materials and the ease of anion formation by SNICS that was exploited in the present study. The source was operated with energy of the $Cs^+$ ions with respect to the negatively biased targets at 5.0 keV. The negative atoms and clusters $C_x^-$, $CsC_x^-$ ( x≥1) were extracted from the source at constant beam energy of 30 keV. Beam energy is defined by the target bias and the extraction voltage. A 30 degree bending magnet with maximum magnetic field of 1 Tesla was used to analyze the anions as a function of m/z. The first mass spectrum was obtained from the pristine samples of MWCNT and graphite at energy of cesium $E(Cs^+)$ = 5.0 keV. The spectra from MWCNTs shown in Fig. 1 contained $C_1^-$, $C_2^-$, $C_3^-$, $C_4^-$ and low intensities of higher clusters as anions. In addition, $O^-$ and $H_2O^-$ were sputtered from the $Cs^+$ bombarded MWCNT cylindrical surfaces and are due to the presence of the adsorbed water.

*2.2. Mass spectrometry of carbon cluster anions versus the cations*

In experiments where one monitors clusters that contain varying numbers of the constituents, anion forming sources may be superior to the cation forming ones. This is due to the possibility of fragmentation and dissociation in charge removing processes that occur in plasma sources. SNICS can be considered an ideal arrangement for experiments to study the sputtered C clusters from nano surfaces like those of the fullerenes, single and multi-walled carbon nanotubes. The high electron affinities of the sputtered C clusters facilitate their detection. Carbon clusters are sputtered predominantly in neutral state and detected subsequently as anions after passing through the neutral $Cs°$ coated surface and the electron cloud in the ionizer. For such experiments the detection of anions has its advantages over that of the cations.

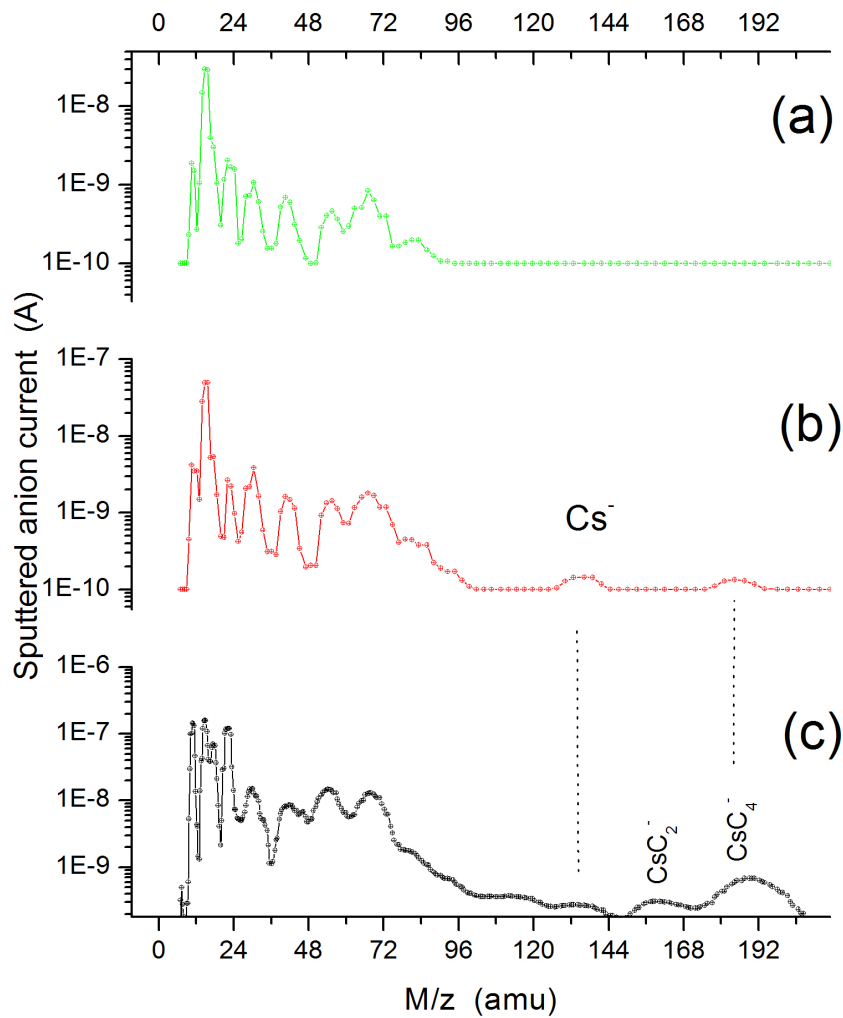

**Fig. 1.** Intensity of the sputtered anions is plotted on the y-axis as a function of the mass to charge ratio (m/z) of the monatomic ($C_1$), diatomic ($C_2$) and higher C clusters ($C_x$; x>2) on the x-axis in atomic mass units (amu). (a) Spectrum from the pristine sample of MWCNTs shows only the carbon anions i.e. $C_1^-$ at 12, $C_2^-$ at 24, $C_3^-$ at 36 amu and the higher ones up to $C_7^-$. The water peak at m/z=18 amu, can also be seen. (b) The spectrum after 20 minutes of irradiations shows similar features with twice the intensity for all of the masses as shown in 1(a) and the m/z peaks for $Cs^-$ and $CsC_4^-$ can be identified. This is the first evidence of cesium coverage of the irradiated nanotube surfaces. (c) The clusters shown have been sputtered from the heavily irradiated ensemble of MWCNTs after 200 minutes of $Cs^+$ bombardments with a cumulative dose of $10^{20}$ ions m$^{-2}$. Two orders of magnitude higher intensities for the sputtered species are emitted as compared with 1(a). Broad humps of the cesium-substituted $CsC_2^-$ and $CsC_4^-$ are clearly visible.

Production of positive charges requires hot plasmas to remove at least one electron from the respective clusters. Our earlier work with the formation of clusters in the hot, carbonaceous environment of the cluster forming soot [12] has shown it to be less efficient and far more complex due to the possibility of fragmentation during excitation and differences in ionization mechanisms for a variety of C clusters. The $Cs^+$ sputtered species are mostly neutrals in the ground state, while the emitted species from plasma sources are invariably in the excited and ionized states. The relatively cooler sputtering environment of SNICS as compared to that of the high temperature carbonaceous discharges has provided us with the entire spectrum of not only the carbon clusters $C_x$ as anions but that of the cesium-substituted clusters $CsC_x$ too. Anion formation with electron attachment preserves the initial bonding configuration of the sputtered clusters.

5 keV $Cs^+$ ion beam was used for this study. $Cs^+$ penetrates ~9 nm depositing its energy in target through direct recoils, creating mono-, di- and tri- vacancies in the sheets or shells and generate collision cascades before being buried under the surface [13]. The accumulation of $Cs^+$ under and $Cs°$ on the surface creates a cesium-mediated carbon cluster sub-surface from which all subsequent sputtering takes place. The neutral cesium $Cs°$ that is adsorbed on the outermost surface further plays its role in donating negative charge to the predominantly neutral, sputtered species and to mediate in the formation of large clusters [10,11].

## 3. MWCNTs

5.0 keV cesium ions $Cs^+$ bombarded the surfaces of the compressed MWCNTs. Fig. 1 has three spectra from $Cs^+$ bombarded surfaces of a large ensemble of 10-20 nm diameter and ~5-15 μm long cylindrical structures that are packed randomly. We packed about $10^8$-$10^{10}$ multi-walled carbon nanotubes in the Al bullet with a surface area ~ 12 $mm^2$. Typical exposed surface area ~ $10^2$-$10^3$ $nm^2$ per nanotubes to the irradiating $Cs^+$ ions. The first mass spectrum in Fig. 1(a) is from a pristine sample that was sputter cleaned for 5 minutes at 5.0 keV. It shows the intensities of the monatomic, diatomic and higher C clusters plotted as a function of respective mass to charge ratio m/z. The intensities are plotted on a logarithmic scale to highlight the emergence of the higher clusters and to ascertain the importance of the nature of the cesiated surface with time.

Higher cluster intensities are order of magnitude less intense therefore, these were plotted on a logarithmic scale to be visualized. The largest sputtered C cluster emitted from the pristine sample is $C_7^-$. In Fig. 1(a) the cluster intensities are low ~ few nA implying the respective number densities ~$10^{10}$ $s^{-1}$. The $O^-$ and $H_2O^-$ sputtered from the adsorbed water are present with order of magnitude higher intensities than the corresponding $C_x^-$ anions. This shows the water adsorbing character of the nano-surface of MWCNTs. It has been difficult to remove by annealing. In Fig. 1(b) the intensities of carbon clusters are 2-3 times higher than those in the pristine sample. This spectrum has the first, clear evidence of the sputtered $Cs^-$, however, with fairly low intensity. Similarly, the first cesium-substituted carbon cluster $CsC_4^-$ is also observed in this spectrum. Evidence of the sputtering of Cs by $Cs^+$ ion beam is important for two reasons: (1) Cs is ejected from the surface by the outward moving constituents of the collision cascades from beneath the surface, (2) Cs° adsorption energies must be much smaller than the binding energies of the C atoms of the MWCNTs. The presence of these two peaks may further help explain some of the conflicting opinions about the presence and the nature of large clusters $C_x$;x>10 that were reported and discussed [4]. At higher $Cs^+$ doses, the other C species, especially $C_1^-$ and $C_2^-$ become comparable to the water constituents that are directly sputtered from the outer surfaces. It can be more effectively removed from graphite surface than from the MWCNTs. That is due to the nature of the vastly different surfaces of the two carbon allotropes. In the last spectrum Fig. 1(c), one can see the still higher intensities ~ 10-100 nA for the anions. $C_1^-$ to $C_6^-$ have well defined structures with $C_7^-$ recognizable on the falling edge of $C_6^-$. $Cs^-$ can still be traced and two well formed peaks are due to $CsC_2^-$ and $CsC_{4-6}^-$. Peaks due to large clusters in all three spectra in Fig. 1 are broad and overlapping.

## 4. Graphite

The anions sputtered from graphite under identical conditions to those of MWCNTs, show similar picture of the Cs mediated C cluster formation on and around the bombarded surface. Fig. 2 has three spectra of C anions with much higher number densities compared with those from MWCNT sputtering. These spectra are taken as a function of the cumulative Cs dose. Fig. 2(a) is from a nascent surface that was sputter cleaned for 5 minutes. Its mass spectrum has all the earlier observed carbon clusters emitted from MWCNTs but with enhanced ~ 10 nA intensity. Higher clusters up to $C_{10}^-$ are present while higher masses (m/z ≥ 133) are unresolved in the tail.

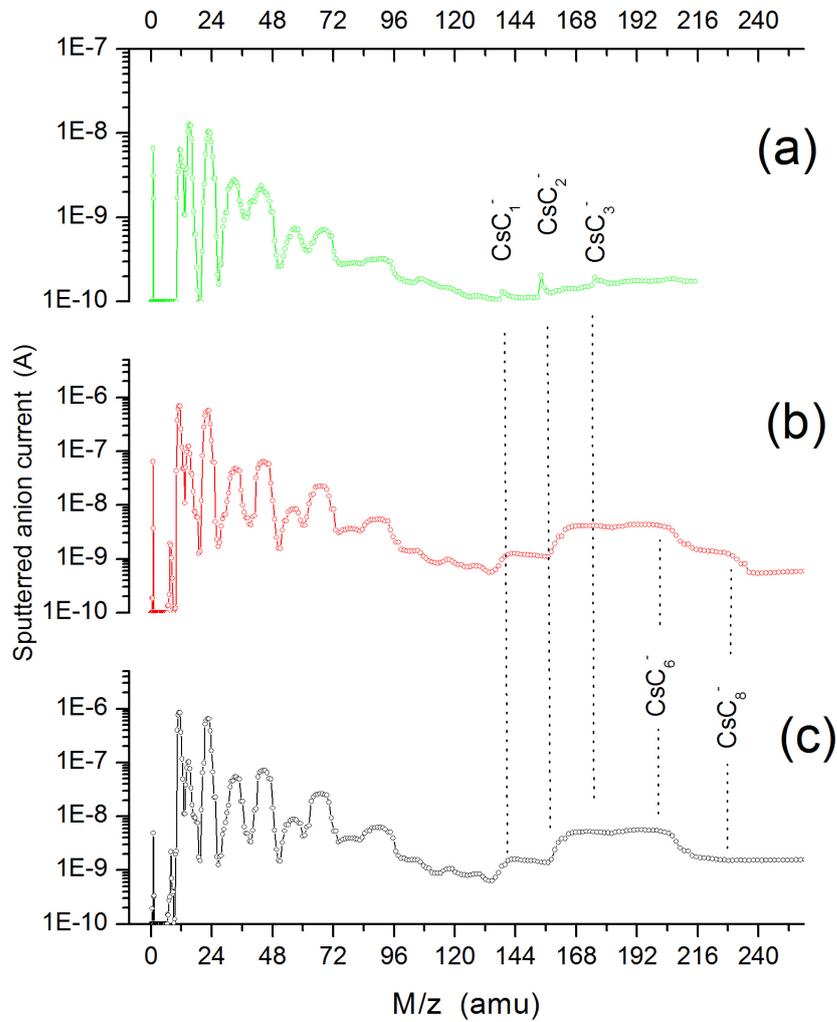

Fig. 2. Three spectra of the intensities of the sputtered species from graphite are plotted as a function of the anion masses (m/z in amu). (a) The sputtered carbon anions from the pristine sample of graphite show the entire range from $C_1^-$ to $C_8^-$ with decreasing intensities. The presence of $CsC_x^-$ (x=1-3) is indicated in the spectrum. (b) After 20 minutes of irradiations two orders of magnitude higher anion intensities can be observed for $C_1^-$, $C_2^-$, $C_3^-$ and higher clusters. The much enhanced twin humps beyond m/z≥10 of $CsC_x^-$ (x=1,2) and $CsC_x^-$ (x=3-6), are also visible. (c) The final spectrum obtained after a cumulative dose of $10^{20}$ $Cs^+$ ions $m^{-2}$. Clusters with even number of C atoms have higher intensities than those with odd ones. Intensities of the anions in 1(c) are in the $10^{-6}$ A range for the smaller anions compared with tens of $10^{-9}$ A from the pristine sample.

The Cs$^-$ peak cannot be clearly resolved among graphite's sputtered clusters. The signatures of CsC$_2^-$ and CsC$_{1-11}$ can be seen, all mixed in broad humps. Intense peaks due to H$^-$, O$^-$ are also present. In Fig. 2(b) all of the anion intensities are higher by at least an order of magnitude. Intensities of the odd ones are lower than those of their next higher, even numbered carbon anions. The cesium-substituted carbon clusters CsC$_1^-$ to CsC$_8^-$ are present in the broad, unresolved peaks. The peak intensities are higher ~ 40 nA. Fig. 2(b) and 2(c) present the spectra from highly bombarded surfaces that have accumulated large quantity of Cs under and near the surface. Two important features in these two spectra can be identified: (1) The even- odd cluster intensities favour the even ones i.e. C$_4^-$, C$_6^-$, C$_8^-$ are more intense than their one down clusters; (2) The CsC$_x^-$ clusters are pronounced and merged. A broad peak from CsC$_2^-$ to CsC$_{11}^-$ is quite visible. This aspect of graphite sputtering is similar but much more enhanced than that from the MWCNT sputtering.

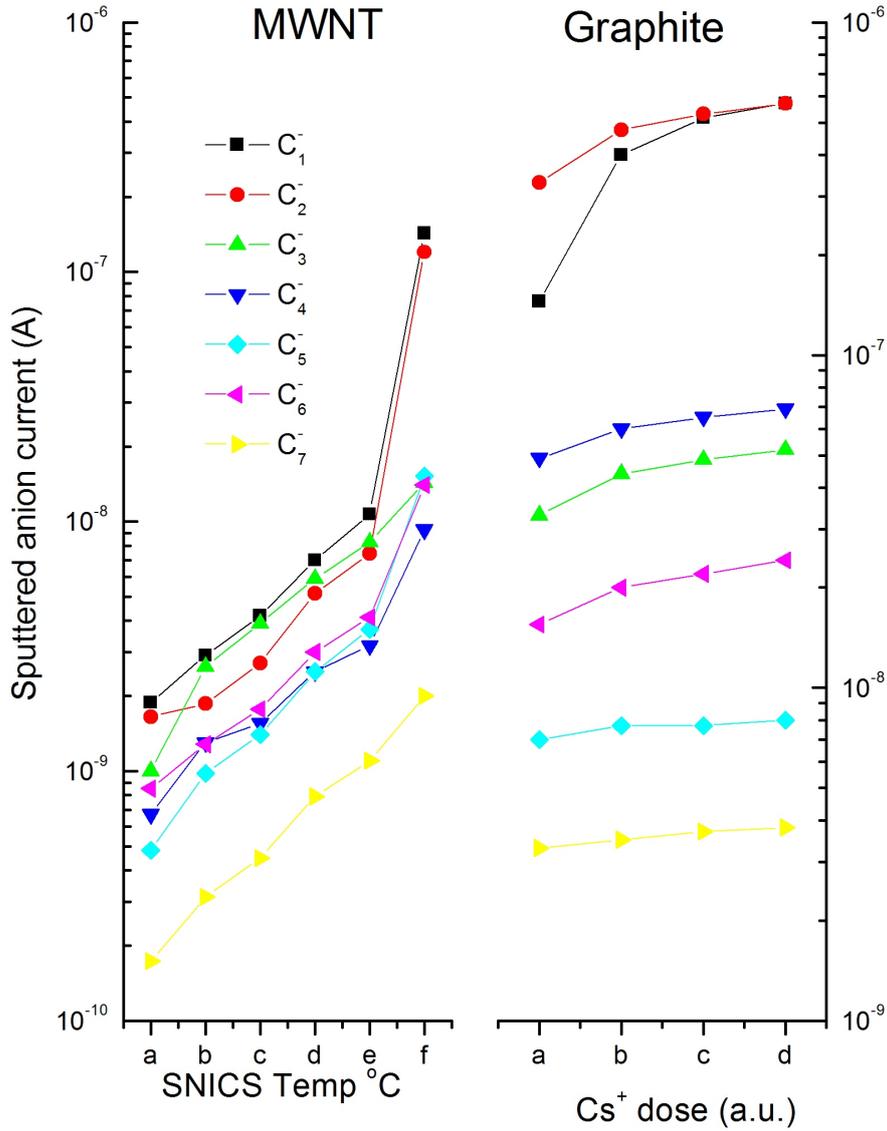

**Fig. 3.** The sputtered anion current densities of the monatomic, diatomic and higher carbon clusters $C_x^-$ (x=1 to 7) are shown for six irradiations from MWCNT as a function of SNICS temperature for the $T_{source}$ = 23°, 45°, 70°, 90° and 105° C, while the last one at *f* is after a cumulative dose of $10^{20}$ $Cs^+$ ions $m^{-2}$. For graphite the same anions $C_x^-$ (x=1 to 7) are shown as a function of the increasing $Cs^+$ dose in four equal steps. The intensity scale for MWCNT sputtered anions is from $10^{-10}$ to $10^{-7}$ Amp while for graphite the range is from $10^{-9}$ to $\sim 10^{-6}$ Amp. Anions intensities show sharp increase as a function of $T_{source}$ for MWCNTs while these show constancy in the case of graphite as a function of the $Cs^+$ dose.

## 5. Role of the adsorbed and implanted Cs on $C_x^-$ output

Fig. 3 plots the sputtered anion intensity under $Cs^+$ bombardment for MWCNT and graphite. In the case of MWCNTs the SNICS temperature is chosen as a parameter. The target current that depends upon Cs vapour density which further related to the source temperature, First five spectra are at $T_{source}$ = 23˚, 45˚, 70˚, 90˚ and 105˚ C, while the last one at *f* is after a cumulative dose of $10^{20}$ $Cs^+$ ions $m^{-2}$. In the case of anions sputtered from MWCNTs, the first entry is at *a*, from the pristine sample. In all later entries there is a gradual increase in the intensities of the sputtered species except in the case of monatomic $C_1^-$ and the diatomic $C_2^-$ that show increase of about an order of magnitude from a highly irradiated MWCNT ensemble. From MWCNTs, one notices that sputtering of clusters shows variations as a function of the Cs˚ coverage of the surface as well as the implanted $Cs^+$ ion density. Both of these facts are connected through the prolonged SNICS operation and the properties of the Cs source. The source temperature acts like an indicator of the $Cs^+$ intensity as well as the target species coverage by Cs˚. Therefore, its impact was seen to have a much larger, surface coverage-related effect in MWCNTs rather than on graphite. In the case of graphite, the Cs˚ coverage gets saturated quickly and we did not observe any significant changes to occur as a function of the SNICS source operation especially the temperature. These two aspects are highlighted in Fig. 3. $C_1^-$ to $C_7^-$ being the major sputtered anions. These were the anions whose intensities could be ascertained within the 20% margin. For the higher species like $C_8^-$, $C_9^-$ and $C_{10}^-$ it is difficult to relate their intensities in all the spectra as a function of the radiation dose, especially from MWCNTs. Graphite on the other hand emits $C_8^-$ that has higher number densities than $C_7^-$. Graphite shows a steady emission for an extended range of $Cs^+$ dose. It shows convincingly that even clusters' output is higher than that of the odd ones. This has been consistently observed by other researchers [3,4,10].

## 6. SEM micrographs of un-irradiated and irradiated MWCNTs and graphite

The surface structure of the un-irradiated and irradiated samples for the two allotropes reveal that the MWCNTs have developed a merged or welded structure after prolonged bombardment by 5.0 keV $Cs^+$. Fig. 4 shows the sample mounted in Al bullet in the centre of the figure at low resolution; almost whole of the sample can be seen. The top micrograph is for the white square on the right, showing the un-irradiated region near the outer radius of the compressed MWCNT sample. At 10,000 times resolution, one can see the bundles of nanotubes with large empty spaces in between. The nanotubes are stacked rather loosely as can be seen.

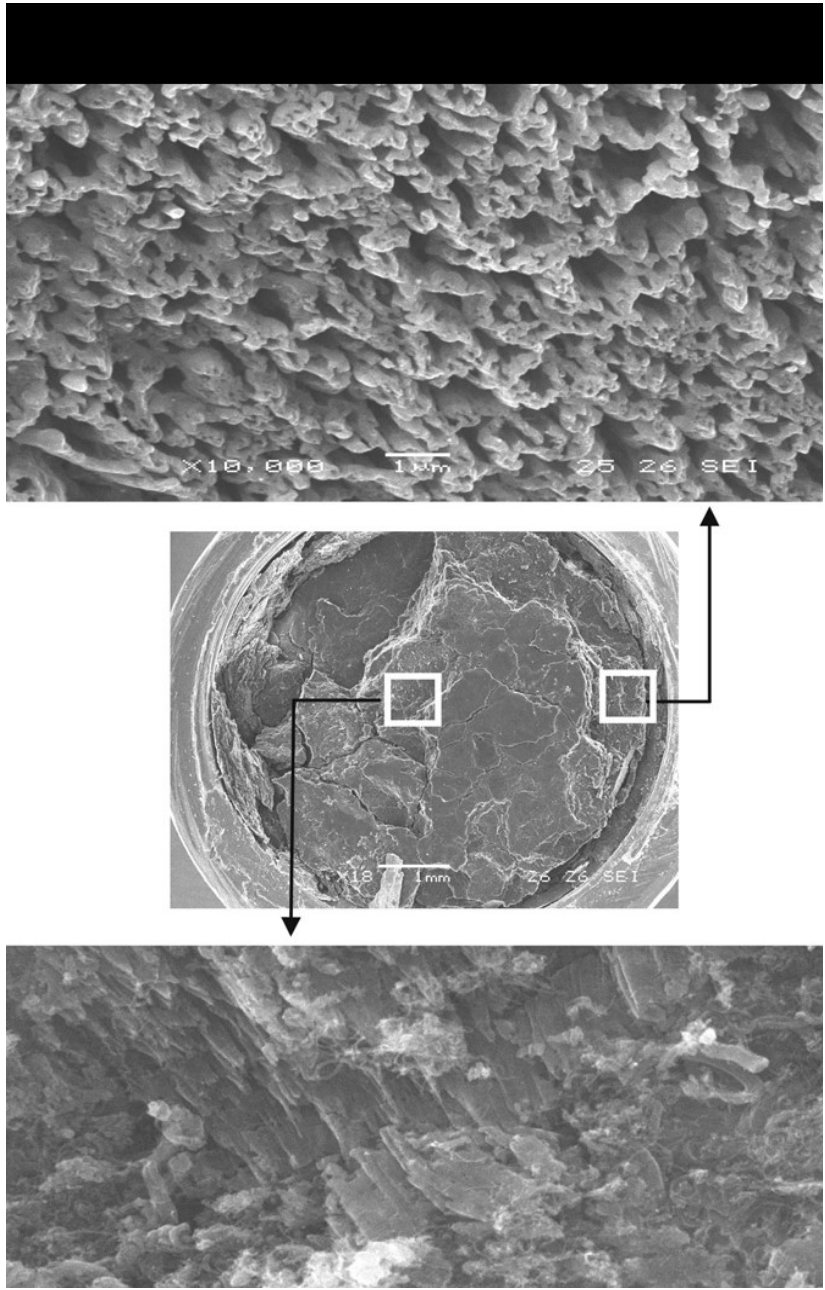

**Fig. 4.** The SEM micrographs of MWCNTs show the un-irradiated portion at the top with x10,000 enlargement and the heavily bombarded region at the bottom with similar magnification (x10,000). The whole sample is shown in the middle with white boxes indicating the micrographed regions. The un-irradiated regions show the separated and recognizable bundles of MWCNTs. The irradiated one shows diffuse and perhaps, partially welded nanotubes.

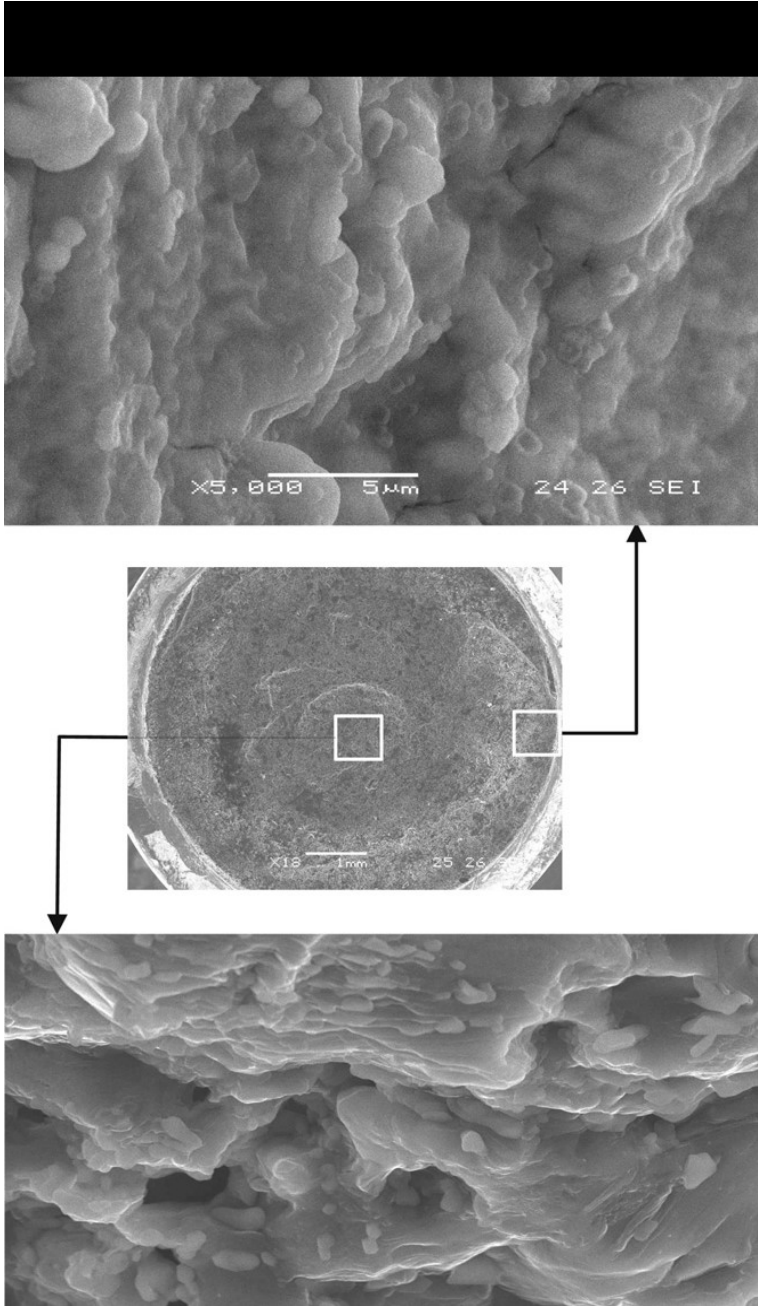

**Fig. 5.** The graphite micrographs for the sample have the entire sample in the center with boxes indicating the un-irradiated region at the top (x5,000) and the irradiated section from the center shown at the bottom (x4,000). No clear differentiation between the two micrographs can be made. Structural features of similar dimensions are seen.

The micrograph shown at the bottom of the figure is for the central region that has been subjected to heavy ion bombardment and hence the associated damage is maximized in this region. One can identify the filling of the open spaces. The tubular bundles are more stacked and the open spaces have been filled. We believe that the cylindrical tubes have been heavily damaged with the subsequent sideways sputtering of the radical carbon atoms and clusters acting as fillers and bonding agents. The situation when compared with that for graphite is not so convincingly clear whether structural changes have occurred or not as shown in Fig. 5. Here again the central picture is that of the graphite sample. The micrograph at the top is for the un-irradiated, edge section at 5,000 times enlargement. The graphite sample is composed of grains that are welded together at the stage of preparation. There are no major structural changes clearly visible in the micrographs of the un-irradiated versus the ion-damaged regions except that the terraces in the lower micrograph are well defined as compared with those at the top, these may be due to the sputter induced step formation.

## 7. Discussion

A clear experimental demonstration of the Cs mediation in cluster formation on the surface is presented in the case of multi-walled carbon nanotubes and graphite. Formation of both the pure carbon clusters $C_x^-$ and the cesium-substituted carbon ones $CsC_x^-$ take place on the surface. The $Cs^-$ peak in the mass spectra from MWCNT identifies the presence of Cs on the surface. It shows this effect to be due to the neutral $Cs°$ on the surface that is being ejected when the collision cascades intersect the outer, $Cs°$-covered surface. The mass spectra show that Cesium seems to mediate the formation of clusters of carbon due to its presence under and on the outer surface of the cylindrical nanotubes and on the exposed surfaces of the grains in graphite. The pattern of $Cs^+$ induced fragmentation has similarities and subtle differences in the two cases. $C_1$s are more numerous than $C_2$s from MWCNTs while the opposite is true in the case of graphite at low doses. This seems to be the case under all irradiation conditions for MWCNTs i.e., the source temperature, $Cs^+$ dose and the state of the structural damage. $C_7^-$ to $C_{10}^-$ are not clearly identifiable from MWCNT spectra while $C_8^-$ to $C_{10}^-$ can be labeled in the spectra from graphite. Among the cesium-substituted carbon clusters, the ones with even number of carbon atoms are favoured in both the target species. $CsC_4^-$ being the first such species to emerge from the MWCNTs. Such observations have been made earlier [10], however, our consecutive spectra from $Cs^+$ bombarded targets, show conclusively, the emerging clusters as a result of the cumulative irradiation effects. We suggest that all clusters are formed on the cesium covered surfaces; MWCNTs providing a much larger landscape for this to happen. Under heavy irradiations, one of the two allotropes i.e. MWCNTs shows some structural transformation. A small change in graphite's average grain size may have occurred but it is not distinctly clear in the micrographs. The sputtered carbon radicals emerge in all directions from the bombarded surfaces and may act as fillers. The individual identity of the heavily bombarded MWCNTs may give way to the merged structures. Similar effects have been reported in the literature on carbon nanotubes irradiations [14-18]. We propose that the re-deposition of the sputtered C clusters may be responsible for the localized melting in MWCNTs leading to the relative absence of the recognizable nanotube bundles in Fig. 4(a) into the observed transformed structures in Fig. 4(b). Randomization of the two allotropic structures might eventually lead to surface amorphization.

## 8. Conclusions

The comparison between the mass spectra of the species sputtered from energetic heavy ion irradiations of the two allotropic forms of carbon MWCNTs and graphite has been reported in the present study. Carbon's monatomic ($C_1$), diatomic ($C_2$), tri-atomic ($C_3$) species and larger clusters $C_x$ (x > 3) are momentum analyzed and detected as sputtered anions from both the allotropic forms. Keeping in view that graphite has been studied over the last few decades in similar sources by a number of researchers, graphite' sputtering pattern provided the standard against which the emissions from MWCNTs can be evaluated. Emissions from the pristine samples from both allotropes show similar order of magnitude intensities for $C_x$ i.e. ~few nano amperes. This implies number densities~$10^{10}$ $s^{-1}$. The heavily irradiated samples show differences in $C_x$ intensities; from MWCNTs the increase ~15 times while graphite shows two orders of magnitude increase. The well quoted result by a number of researchers [3,4,10] is that cluster intensities for the even ones are higher than the preceding odd ones; we also confirm this and can be seen in Fig. 2(c). The situation is different from MWCNTs where a comparison of cluster intensities I($C_x$) shows that I($C_3$)>I($C_4$) and I($C_5$)≈I($C_6$) as can be seen in Fig. 1(c). The effects of the large quantities of sputtered species and their accumulation in the intra-MWCNT space might be responsible to the merged and perhaps, welded-looking structures in the micrographs. We conclude that highly irradiated carbon nano-structures are internally massively deformed and the sputtered species $C_x$ may accumulate and form connecting bridges between the nanotubes. In graphite this may not be the case for being a uniformly distributed; μm sized grain-based macro-structure.